\documentclass[pra,aps,twocolumn,showpacs,superscriptaddress,longbibliography]{revtex4-2}

\usepackage{ulem}
\usepackage{graphicx}
\usepackage{amsfonts}
\usepackage{epstopdf}
\usepackage{color}
\usepackage[dvipsnames,svgnames]{xcolor} 

\begin{document}

\title{Observation of the noise-driven thermalization of the Fermi-Pasta-Ulam-Tsingou recurrence in optical fibers}

\author{Guillaume Vanderhaegen}
\affiliation{University of Lille, CNRS, UMR 8523-PhLAM-Physique des Lasers Atomes et Mol\'ecules, F-59000 Lille, France}
\author{Pascal Szriftgiser}
\affiliation{University of Lille, CNRS, UMR 8523-PhLAM-Physique des Lasers Atomes et Mol\'ecules, F-59000 Lille, France}
\author{Alexandre Kudlinski}
\affiliation{University of Lille, CNRS, UMR 8523-PhLAM-Physique des Lasers Atomes et Mol\'ecules, F-59000 Lille, France}
\author{Matteo Conforti}
\affiliation{University of Lille, CNRS, UMR 8523-PhLAM-Physique des Lasers Atomes et Mol\'ecules, F-59000 Lille, France}
\author{Andrea Armaroli}
\affiliation{University of Lille, CNRS, UMR 8523-PhLAM-Physique des Lasers Atomes et Mol\'ecules, F-59000 Lille, France}
\author{Arnaud Mussot}
\affiliation{University of Lille, CNRS, UMR 8523-PhLAM-Physique des Lasers Atomes et Mol\'ecules, F-59000 Lille, France}

\begin{abstract}
We report the observation of the thermalization of the Fermi-Pasta-Ulam-Tsingou recurrence process in optical fibers. We show the transition from a reversible regime to an irreversible one, revealing a spectrally thermalized state. To do so, we actively compensate the fiber loss to make the observation of several recurrences possible. We inject into the fiber a combination of three coherent continuous waves, which we call Fourier modes, and a random noise. We enhance the noise-driven modulation instability process against the coherent one by boosting the input noise power level to speed up the evolution to the thermalization. The distributions of the Fourier modes power along the fiber length are recorded thanks to a multi-heterodyne time-domain reflectometer. At low input noise levels, we observe up to four recurrences. Whereas, at higher noise levels, the Fourier modes fade into the noise-driven modulation instability spectrum revealing that the process reached an irreversible thermalized state.
\end{abstract} 

\date{\today} 
\maketitle
\section{Introduction}
The study of the Fermi Pasta Ulam Tsingou (FPUT) paradox in the 50s \cite{Fermi_1955} marked the birth of numerical simulations and of nonlinear physics \cite{FPUmath}. These numerical experiments aimed at studing the energy transfers between the eigenmodes of a chain of oscillators. A linear coupling makes the excited mode preserving its energy, while a nonlinear coupling enables energy exchanges. Fermi and co-workers were expecting an equidistribution of the energy into all the modes. But, against all odds, the system exhibited a reversible behavior, and, after a certain time, returned to its initial state.

This recursive energy transfer is rather general and can be observed in many other fields of physics, where a weak nonlinear coupling exists between the system eigenmodes. It is now conventionally denoted as FPUT recurrence process. Such behaviors were especially investigated in focusing cubic media through the nonlinear stage of modulation instability (MI), as demonstrated in hydrodynamics \cite{Lake_1977, Kimmoun_2016}, bulk crystals optics \cite{Pierangeli_2018}, planar waveguides \cite{Cambournac_2002}, fiber optics \cite{VanSimaeys_2001, Van_Simaeys_2002, Hammani_2011, Bendahmane_2015, Mussot_2018, Hu_2018, Goossens_2019, Vanderhaegen_2020} and also magnetic feedback rings \cite{Wu_2007}. A recurrent regime is obtained by triggering these systems with a weak coherent seed aside the pump located inside the MI gain band. During the propagation, the pump transfers energy to this input signal and to the symmetric idler wave generated by four wave mixing, the Fourier modes of the systems. Higher order Fourier modes, also called harmonics, are then generated by four-wave mixing process up to a specific distance before the energy transfers reversed back and the system returns to the initial excitation state \cite{VanSimaeys_2001, Van_Simaeys_2002, Hammani_2011, Bendahmane_2015, Mussot_2018, Hu_2018, Goossens_2019, Vanderhaegen_2020}.

A major question was to figure out if these systems beyond a single recurrence period could reach a thermalized state, as firstly expected by Fermi, Pasta, Ulam and Tsingou. The limited performances of their MANIAC computer \cite{FPUmath} didn't allow to pursue the study until this irreversible final state. Indeed, the thermalization is a long-process of energy equipartition between the eigenmodes of the mechanical oscillators requiring lengthy calculations. Fortunately, the computational power of modern computers enabled the achievement of these complex calculations. Hence, they revealed that after several recurrences, pseudo-recurrences appear before the system eventually reaches a thermalized state \cite{Lvov2018, Benettin_2013, Onorato_2015, Pace_19}. By analogy, in fiber optics, the irreversible evolution towards an energy distribution among the whole spectrum (not only harmonics of the signal, but any spectral components including noisy ones) after the recurrences breakdown, was called thermalization too \cite{Wabnitz_2014}. In that case, the noise-driven MI process enters into competition with the seeded one, which becomes eventually the dominant effect \cite{Wabnitz_2014}. Aside the FPUT recurrence process, this competition was deeply investigated in the context of supercontinuum generation and rogue waves because MI is the triggering mechanism of these complex nonlinear processes \cite{Akhmediev_2009, Dudley_2014, Toenger_2015, Suret_2016, Narhi_2016, Kraych_2019, Gelash_2019, Dudley_2009}. A coherent seed was used to either increase supercontinuum coherence or inhibit rogue wave formation \cite{Solli_2008,Dudley_2008}. 

While the nonlinear stage of MI through the prism of FPUT recurrences in either noise-driven systems \cite{Narhi_2016, Kraych_2019} or coherently seeded ones \cite{VanSimaeys_2001, Van_Simaeys_2002, Hammani_2011, Bendahmane_2015,Hu_2018,Mussot_2018,Goossens_2019} had already been studied, their competition and the route to a thermalized state had never been reported experimentally. Indeed, for a while, the recurrence number was limited to a single cycle \cite{VanSimaeys_2001, Van_Simaeys_2002, Hammani_2011, Bendahmane_2015, Hu_2018} or two periods \cite{Mussot_2018,Goossens_2019} due to the fiber loss. Indeed dissipation, while weak in optical fibers, dramatically increases the period of recurrence. This prevents the study of the thermalization process that is expected to appear after many periods \cite{Lvov2018, Benettin_2013, Onorato_2015}. Recently, it was possible to overcome this issue in fiber optics thanks to an active loss compensation scheme through Raman amplification \cite{Mussot_2018, Vanderhaegen_2020, Naveau_2021} or the use of a recirculating loop \cite{Kraych_2019}. 

In this work, we investigate the competition between noise-driven and seeded MI processes in the framework of multiple FPUT recurrences to report the observation of the thermalization process. As in Refs. \cite{Mussot_2018, Vanderhaegen_2020, Naveau_2021}, we compensate almost perfectly the loss through Raman amplification to get several recurrences periods and we accelerate the route to thermalization by increasing the noise power level at the fiber input. It enables to finely control the balance between both MI processes to clearly observe the transition from a perfectly recurrent regime to an irreversible spectrally thermalized state. The paper is organized as follows. First, we provide numerical simulations to illustrate the noise and coherently driven MI processes and an example of competition between both. Second, we introduce our experimental setup, made of a multi-heterodyne optical time domain reflectometer (HODTR) system combined with an active loss compensation scheme. Finally, the experimental results are displayed, compared to the NLSE numerics.

\section{Numerical simulation of spontaneous and seeded MI}
\begin{figure}[!h]
\includegraphics[width=1\columnwidth]{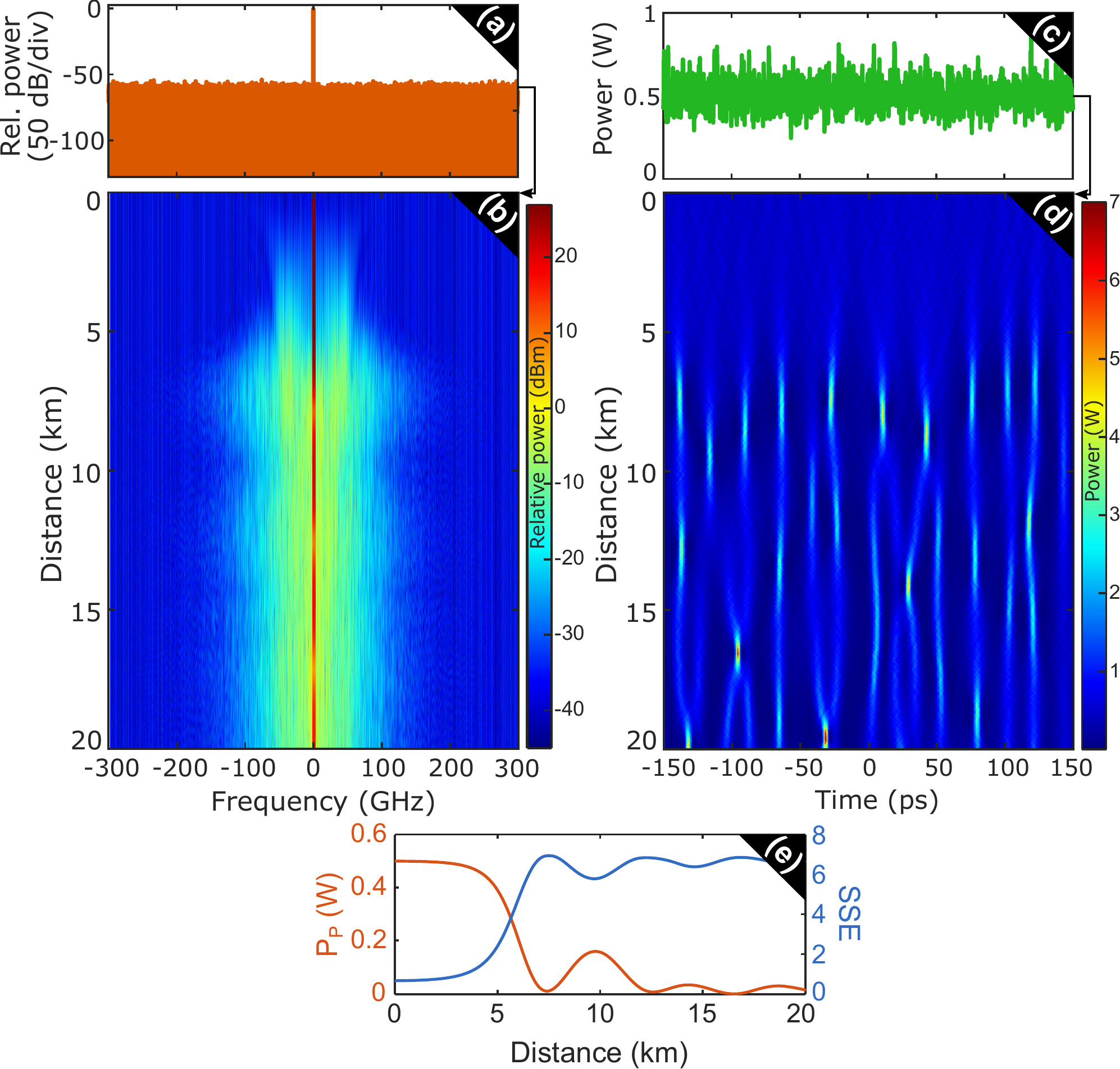}
\caption{Numerical simulations of the dynamic of spontaneous MI. (a) Spectrum at the fiber input, and (b) power spectral distribution over the whole fiber length. (c) Power time profiles at the fiber input, and (d) temporal power distribution along the fiber length. (e) Shannon spectral entropy SSE and power of the initial pump excitation as function of the fiber distance. Parameters: group velocity dispersion $\beta_2=-21$ ps$^2$km$^{-1}$, nonlinear coefficient $\gamma=1.3$ W$^{-1}$km$^{-1}$, input pump power $P_{P}=500$ mW, noise $PSD=-112$ dBm/Hz.}
\label{spont}
\end{figure}

Numerical simulations rely on the integration of the nonlinear Schrödinger equation (NLSE):
\begin{equation}
    \frac{\partial E}{\partial z}=-\frac{\beta_2}{2}\frac{\partial^2 E}{T^2}+i\gamma |E|^2 E
\end{equation}
where $E$ is the electrical field, $z$ is the distance along the fiber and $T$ is the retarded time of the frame moving at the group velocity. $\beta_2$ is the group velocity dispersion with $\beta_2=-21$ ps$^2$km$^{-1}$ and $\gamma$ is the nonlinear coefficient with $\gamma=1.3$ W$^{-1}$km$^{-1}$. This equation is solved numerically using the split-step Fourier method \cite{Agrawal_2007}.
The evolution of the noise-driven MI process as a function of the fiber length is illustrated in Fig.~\ref{spont} with a typical example.
The input spectrum is shown in Fig.~\ref{spont}. (a). A monochromatic pump is surrounded by a white noise floor  with a power spectral density $PSD=-112$ dBm/Hz. In all the numerical simulation, the real and the imaginary part of the input noise spectrum at each frequency bin are samples of two independent Gaussian random variables with zero mean and unit variance. The noise floor is then adjusted at the desired level by multiplying the amplitude of the noise bins with a constant factor.


The evolution of the spectrum along the propagation distance is plotted in Fig.~\ref{spont}. (b) for a single shot of noisy initial condition. At the beginning (up to $5$ km), only the spectral components within the MI gain band (the cutoff frequency is $\omega_C=\sqrt{4\gamma P_P /|\beta_{2}|}$ \cite{Agrawal_2007}) are amplified. Then the spectrum broadens due to four-wave mixing (FWM) between the unstable frequencies. Looking at the input time profile (Fig.~\ref{spont}. (c)) and its evolution along the fiber length (Fig.~\ref{spont}. (d)), we notice the irregular appearance of high power pulses. We can even distinguish solitonic structures similar to the Akhmediev breather (AB), the Kuznetsov-Ma soliton (KM) and the Peregrine soliton (PS) as previously reported in \cite{Suret_2016, Narhi_2016, Kraych_2019}. To highlight the continuous spectral broadening during the propagation, we calculate the Shannon spectral entropy (SSE) \cite{Shannon_1948}:
\begin{equation}
    SSE=- \sum_{\omega} p_{\omega} log_{e}(p_{\omega}) 
\end{equation}
where $p_{\omega}$ is the fraction of total power contained in the frequency bin $\omega$. The evolution of the SSE (solid blue line) together with the pump power along the fiber length (solid orange line) are plotted in Fig.~\ref{spont}. (e). The SSE is initially close to 1 and increases from about 5 km too, when spectral components are generated outside the MI band, to reach a maximum value around 7. Then the entropy slightly oscillates around this maximum during the rest of the propagation. After dropping to almost zero, the pump power oscillates too, which is a signature of spontaneous FPUT recurrences \cite{Kraych_2019}. We notice the SSE and the pump power are antiphased. This shows the system entropy is maximum when the pump power is minimum and vice versa.

\begin{figure}[!h]
\includegraphics[width=1\columnwidth]{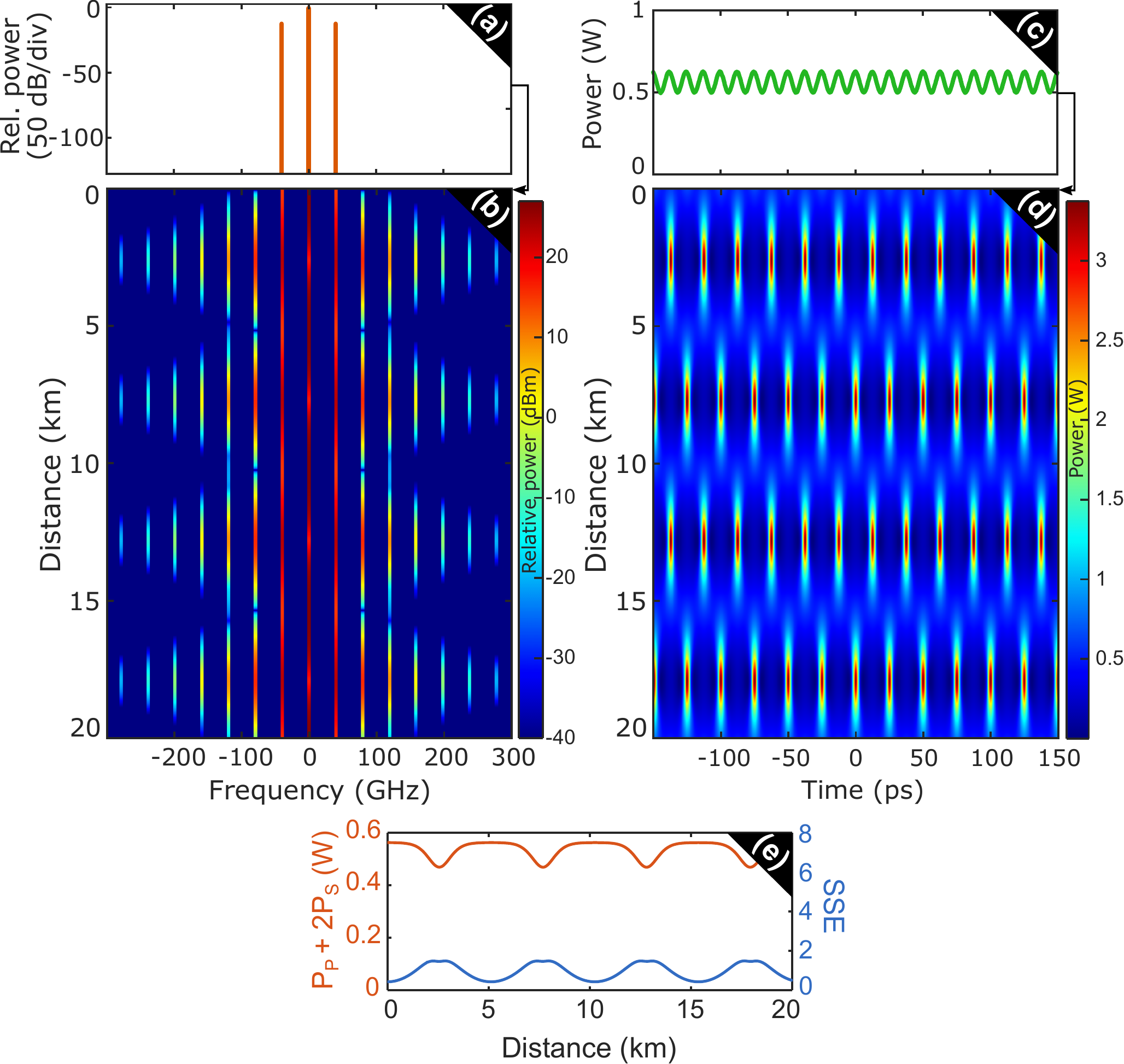}
\caption{Same as Fig.~\ref{spont} but with seededed MI. The modulation wave is set at $12$ dB below the pump wave. On (e) is plotted the total power of initial 3 waves excitation. $PSD \approx -400$ dBm/Hz (numerical noise level).}
\label{seed}
\end{figure}

A numerical simulation of coherently driven MI is presented in Fig.~\ref{seed}. 
We used a set of three waves, a pump and two symmetric sidebands located close to the maximum of the MI gain band \cite{Agrawal_2007} (Fig.~\ref{seed}. (a)). Fig.~\ref{seed}. (b) shows the evolution of the spectrum along the propagation distance. On a first stage, the seeds are amplified by the pump and the spectrum is broadened with equidistant peaks into a triangular shaped frequency comb, at the expense of the pump wave. New waves located outside the gain band, called harmonics, are generated outside the gain band due to multiple four-wave mixing (FWM) processes. Then the system reaches the maximum pump saturation stage around $2.5$ km and the energy transfer to the higher order modes is not possible anymore. Therefore, the energy transfer direction reverses and flows back to the pump wave before the system returns almost exactly to its initial state. This corresponds to a first FPUT recurrence cycle. Then, this dynamics repeats itself periodically during the propagation till the end of the fiber. In this case, we can distinguish 4 recurrences. The temporal evolution along the fiber length in the time domain is depicted in Fig.~\ref{seed}. (d). The system is triggered with a weakly modulated CW (Fig.~\ref{seed}. (c)). We notice the emergence of maximum compression points coinciding with the spectral broadening maxima, whose positions are accurately predicted by the theory \cite{Grinevich_2018, Naveau_2019}. The $\pi$-shift between two consecutive recurrences is due to the initial phase relation between the initial three waves \cite{Mussot_2018, Vanderhaegen_2020_2}. As with spontaneous MI, we plot the SSE of the system in Fig.~\ref{seed}. (e) (solid blue line). The entropy is weak ($<2$), oscillates periodically, at the the FPUT period but doesn't expand. The power contained in the initial waves (orange solid line), the pump and the modulation sidebands, is also periodic. The minima of this curve corresponds to maximal spectral broadenings because a significant part of the total power is contained in the Fourier modes of order $n>2$. 

\section{Numerical simulation of noise-induced thermalization}

\begin{figure}[!h]
\includegraphics[width=1\columnwidth]{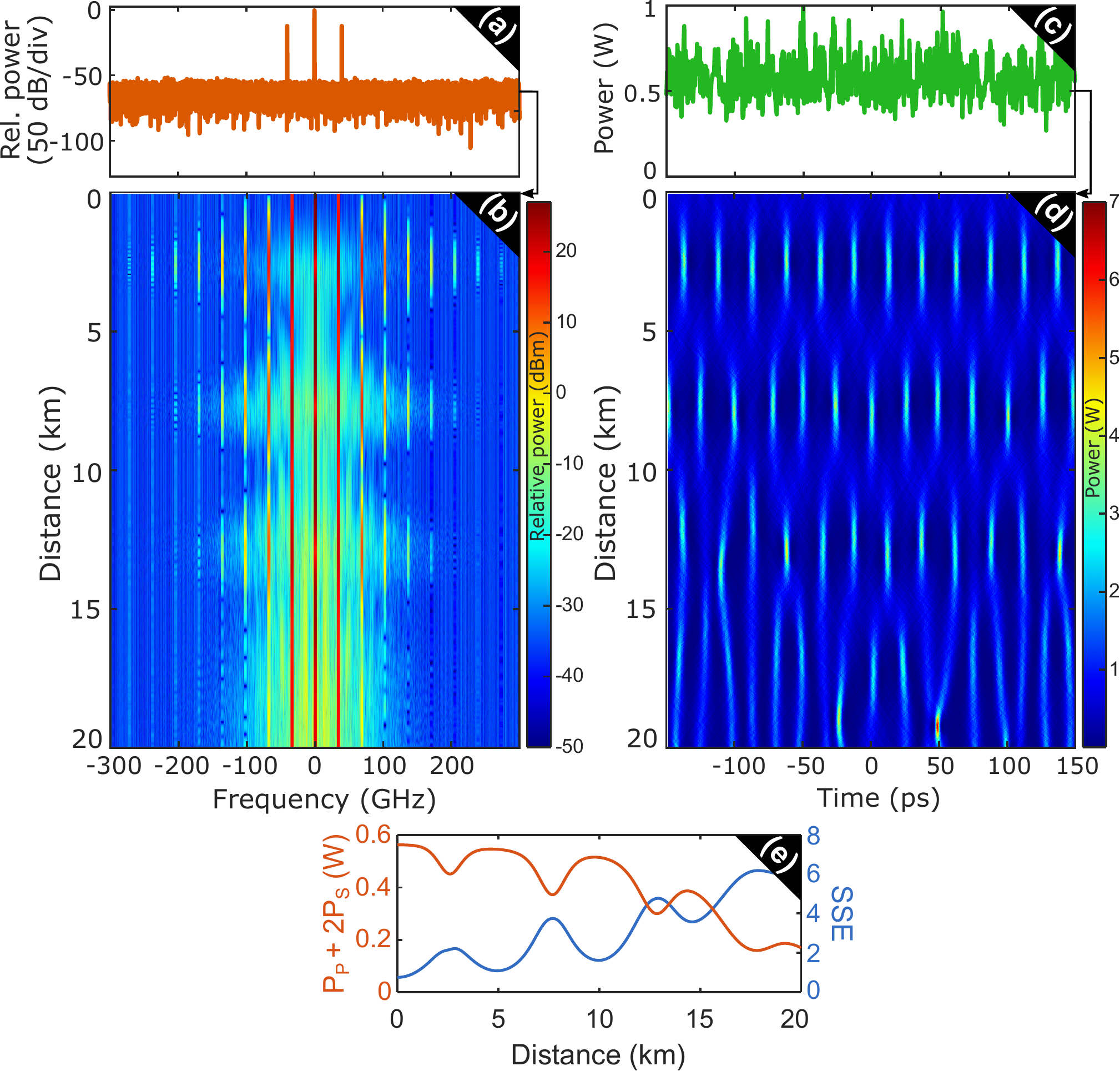}
\caption{Same as in Fig.~\ref{spont} and Fig.~\ref{seed}. Same 3 waves input parameters as the seeded MI numerics but with the spontaneous MI numerics noise floor, $PSD=-112$ dBm/Hz.}
\label{noise_therm}
\end{figure}

We now look at the competition between spontaneous and seeded MI, when the CW is initially modulated by a coherent sinusoidal modulation with noise (Figs.~\ref{noise_therm}. (a) and (c)). 
The spectral and time evolutions are presented in Figs.~\ref{noise_therm}. (b) and (d) respectively. At the beginning of the propagation in the fiber, the dynamics is very similar to the one illustrated Fig.~\ref{seed} with two clear recurrences. This means the dynamics of the system is dominated by the coherently driven MI process. However, by further propagating within the fiber, the noise level of spectral component located within the MI gain band increases and the noise driven MI becomes significant at the expense of the Fourier modes. Higher Fourier orders progressively disappear into the noise floor. The dynamics is then similar to the one illustrated in Fig.~\ref{spont}, with the appearance of irregular temporal pulses and the disappearance of the fourth recurrence. The evolutions of the SSE and the total power contained in the central 3 waves (Fig.~\ref{noise_therm}. (e)) also reveal this transition toward an irreversible state. The SSE keeps increasing to 7, and the total power tends to zero, as for the spontaneous MI example (Fig.~\ref{spont}). A fully thermalized state showing a complete disappearance of the coherent waves could be asymptotically reached by propagating in a much longer fiber (not shown here for the sake of clarity). The transition speed from spontaneous and noise driven MI can be controlled by tuning the input noise level. We will exploit experimentally this feature for the observation of the of the system evolving toward an irreversible thermalized state. 

\section{Experimental setup}
\begin{figure}[!h]
\includegraphics[width=1\columnwidth]{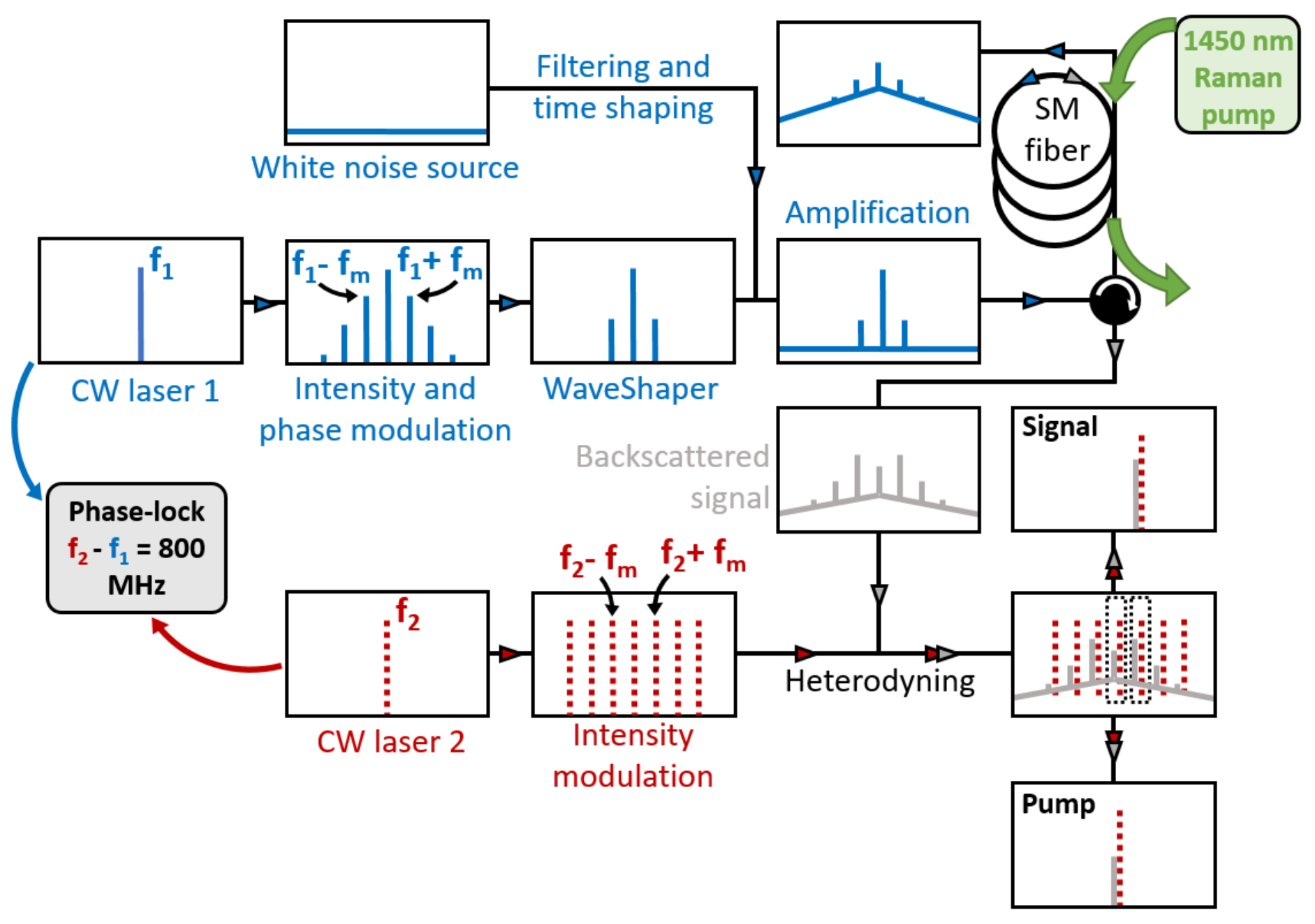}
\caption{Simplified sketch of the setup with "spectrum boxes". $f_{1}$ is the frequency of the main laser and $f_{2}$ of the local oscillator, phase-locked at $800$ MHz from $f_{1}$. $f_{m}$ is the modulation frequency, set here at $38.2$ GHz. SM fiber: single mode fiber. Waveshaper: amplitude and phase programmable optical filter.}
\label{setup}
\end{figure}

The setup is similar to the one used in \cite{Mussot_2018, Naveau_2021} and a simplified version is presented in Fig.~\ref{setup}.
A $1550$ nm CW signal is generated from a single frequency laser diode. To excite the modulation instability regime coherently, the signal is first phase-modulated at the frequency $f_m=38.2$ GHz, very close of the maximum gain frequency ($f_{max}=38.4$ GHz). The signal is then time shaped with an intensity modulator into $50$ ns square pulses, which are short enough to avoid stimulated Brillouin scattering and allow optical time domain reflectometry (OTDR). These pulses are long enough to be considered as a CW wave for the modulation instability time scale. To shape the signal into the desired 3 waves input, the relative power between the Fourier modes are tuned with a Waveshaper. The signal is then combined with synchronized additional white noise pulses, from a power tunable white noise source. It is followed by an amplification stage through an erbium doped fiber amplifier (EDFA) to tune the nonlinear length ($P_P=470$ mW). An acousto-optic modulator, which has a extinction ratio much higher than electro-optic modulators, is then cascaded to reject the inter-pulses noise and to control the pulses sequence. The signal is injected into a $16.8$ km long single mode fiber (G-652) in which nonlinear interactions will occur. 

Even if the dissipation value is low ($\alpha=0.2$ dB/km), on tens of kilometers, the nonlinear dynamic is highly impacted \cite{Naveau_2021}. In order to keep the signal power almost constant over the whole fiber length, allowing us to preserve the integrable NLSE dynamics, an active loss compensation scheme was implemented. At the fiber output, a backward $1450$ nm laser is injected to amplify the signal by Raman effect. The Raman pump is located $13.2$ THz away from the amplified signal to benefit from the maximum Raman gain \cite{Agrawal_2007}. For these tens of ns duration pulses, we can neglect the saturation of the amplifier and consider flat amplified pulses \cite{Vanderhaegen_2021}. 

The power distribution along the fiber of the main Fourier modes are obtained by an OTDR combined with a multi-heterodyne technique \cite{Mussot_2018, Naveau_2021}. At each location in the fiber, a tiny part of the signal is Rayleigh backscattered and combined with a frequency comb with repetition frequency $f_m$, shifted by $600$ MHz from the main laser (the AO modulator induces a $200$ MHz shift on $f_{1}$). The independant recording of the beating between the main Fourier modes allows to recover the relative evolution of the power along the all fiber length. 

The noise power is increased progressively to study the influence of the input noise level on the thermalization of the FPUT recurrences. For each initial noise level value, we record the power distribution of the two main Fourier waves, the pump and the signal (the initial excitation being symmetric, we consider that the idler wave evolves identically to the signal) and the spectrum at the fiber end. Much details about the experimental setup can be found in Ref. \cite{Naveau_2021}.

\section{Experimental results}
\begin{figure*}[ht]
\includegraphics[width=1\textwidth]{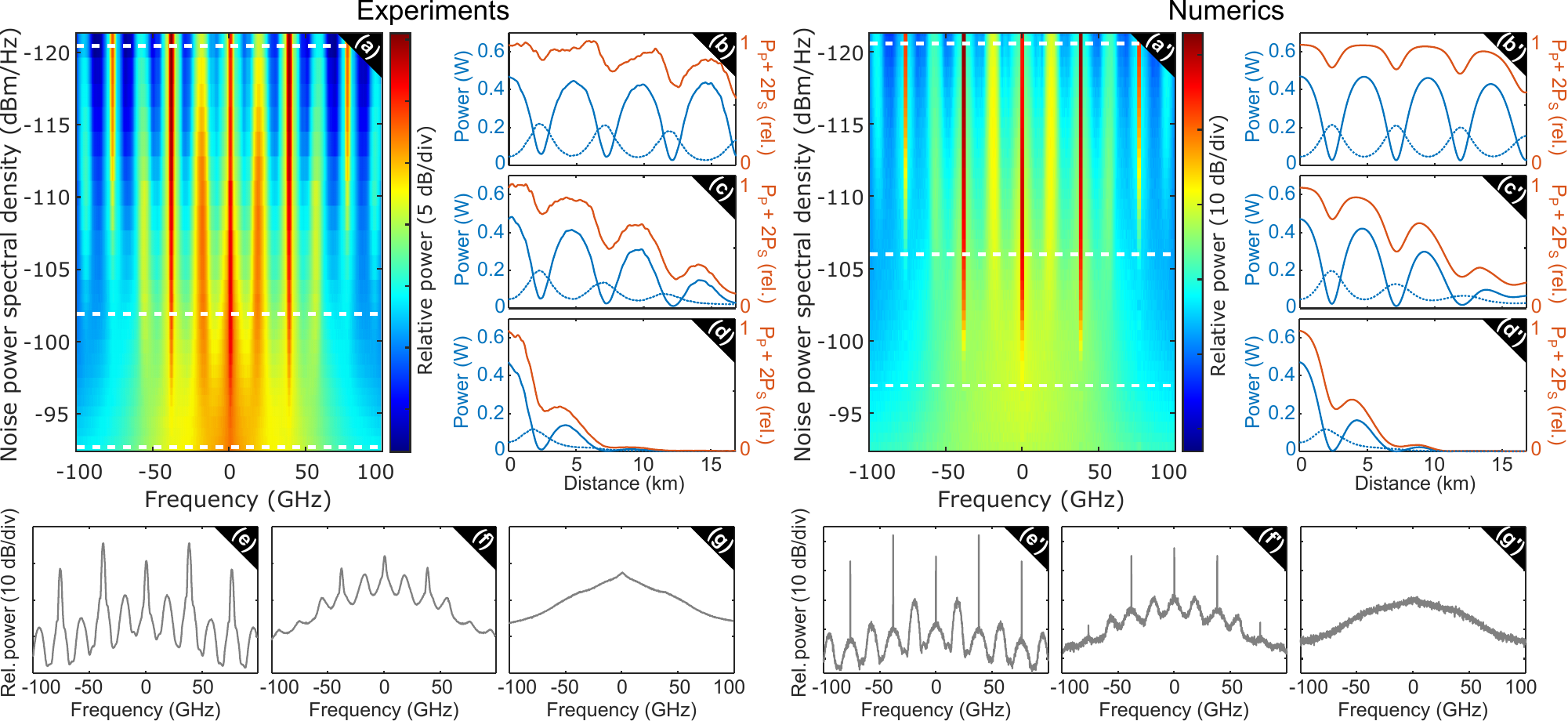}
\caption{(a), (a') Spectrum at the fiber end as a function of the initial power spectral density of noise. (b), (c), (d) and (b'), (c'), (d') Evolution of the pump (solid blue line), the signal (dotted blue line) and the sum of the central 3 waves $P_P+2P_S$ (solid orange line) along the fiber distance. (e), (f), (g) and (e'), (f'), (g') Output spectra. The initial power spectral densities are respectively $(-121.3, -101.8, -92.4)$ dBm/Hz and $(-121.3, -106.4, -97.3)$ dBm/Hz. The left panel is corresponding to the experimental recordings while the right panel to the numerics.}
\label{drawing8}
\end{figure*}

We varied the initial noise power spectral density (PSD) from $-121.3$ to $-91.9$ dBm/Hz. We recorded the pump and signal waves power evolutions along the fiber length as well as the output spectra. These results are presented on the left panel of Fig.~\ref{drawing8}. (a-g). 
Fig.~\ref{drawing8}. (a) displays the output spectra as a function of the initial noise PSD. At low amplitude noise floors, we clearly discern the Fourier modes: the pump, the seeds (at $\pm 38.2$ GHz) and the first harmonics (at $\pm 76.4$ GHz). Between these sidebands, MI also amplifies noisy components. We note spectral holes around signal and idler waves. This is due to multiple four-wave mixing processes involving noise components and these waves \cite{Inoue_2001} (see Fig.~\ref{drawing8}. (e)). By increasing the input noise PSD, the higher order harmonics intensity decreases at the expense of the inter-bands lobes. The first harmonics (at $\pm 76.4$ GHz) are even not visible anymore from the amplified noise floor from about $-105$ dBm/Hz. The seed (at $\pm 38.2$ GHz) are also lost in the noise around $-95$ dBm/Hz. At the same time, the amplitude of the noise floor at the fiber output increases with the initial noise level. Indeed, with a higher input noise level, the pump saturates at a shorter fiber length. This happens at the expense of the coherent dynamic process of seeded MI. The pump is not powerful enough to pursue the power transfer with the other Fourier modes anymore, the spontaneous MI process is taking the lead. 

In Figs.~\ref{drawing8}. (b-d) are represented the pump (solid blue line), the seed (dotted blue line) and the relative total power of the three central waves distributions along the fiber length for noise PSD of $-121.3$, $-101.8$ and $-92.4$ dBm/Hz respectively. For the first case (PSD = $-121.3$ dBm/Hz, see Fig.~\ref{drawing8} (b)), no additional noise has been added and the noise level is the intrinsic noise of the setup (lowest value). We observe $3.5$ recurrences with a very small decrease of the 3 waves total power, as it has been observed in \cite{Vanderhaegen_2020}. This process is then still highly coherent and the seeded MI is dominating the nonlinear power transfers. When the initial noise floor is increased at $-101.8$ dBm/Hz (Fig.~\ref{drawing8} (c)), the power of the Fourier modes decreases during the propagation. It indicates that a part of the power is located elsewhere in the spectrum, mainly in between coherent components. For instance at $z=7.4$ km, (after 1.5 recurrences), only half of the power is located in the Fourier modes and it reaches $17 \%$ at the fiber end. We see in the output spectrum depicted in Fig.~\ref{drawing8} (f) that only the three main Fourier modes remain visible. The seeded MI process is not longer effective and the spontaneous MI is not negligible anymore. In the last case, with a noise PSD of $-92.4$ dBm/Hz (Fig.~\ref{drawing8} (d)), we can only distinguish one and half recurrences and the power of the central three waves becomes lower than $50 \%$  before the system reaches the first half recurrence. After one and half recurrence (about 5 km), the power of both the pump and the seed waves drops drastically during the propagation to reach weak values, comparable to the noise floor. At the fiber output, the spectrum in Fig.~\ref{drawing8}. (g) has a triangular shape from which it is impossible to distinguish any coherent Fourier modes. This shape is typical from parametrically driven systems \cite{akhmediev_universal_2011} and is the signature of the thermalization of the FPUT process in optical fibers. With this high noise input level, all coherent states are hidden in the noise components including the pump. The consequence is that a comeback to a coherent initial state is impossible, the evolution being thus irreversible. This unambiguously proves the system reached a thermalized state of the FPUT recurrence process in optical fibers. 

The results from numerical simulations are plotted in the right panel of Fig.~\ref{drawing8}. 
The agreement between experiments and numerics is very good. The experiments follow the same dynamics than the one predicted through numerics, the progressive coherence loss with the increase of the input noise. Remarkably, the disappearance of coherent structures and the triangular shape of the output spectra for high input noise levels is very similar to experiments (Fig.~\ref{drawing8} (g) and (g')). This confirms the system reached an irreversible thermalized state. 

\section{Conclusion}
We reported the experimental observation of the thermalization of the FPUT process in optical fibers. We showed that the FPUT recurrences irreversibly disappear due to a competition between noise driven and coherently driven MI processes leading to an increase of the system entropy. Our experiments were realized by means of a multi-HOTDR system to monitor the evolution of the power along the fiber length. The observation of the route to the thermalisation is made possible with an active loss compensation scheme to observe up to 4 recurrences and by adding noise at the fiber input to accelerate the process. At high input noise level, we showed that the system reaches an irreversible thermalized state, with an output spectrum only made of noisy components. Experimental results were confirmed by numerical simulations with an excellent agreement. These experimental works revealed the evolution of a nonlinear system from a coherent dynamics between its nonlinear modes toward an irreversible thermalized state. This behavior, originally expected by Fermi, Pasta, Ulam and Tsingou in a nonlinearly coupled masses system, had never been reported experimentally so far. We anticipate these results to be generalizable to all focusing cubic media and to contribute to a better understanding of the dynamics of noisy and coherently driven nonlinear systems and the formation of rogue waves.

\section*{Acknowledgments}
The present research was supported by the agence Nationale de la Recherche (Programme Investissements d’Avenir, I-SITE VERIFICO); Ministry of Higher Education and Research; Hauts de France Council; European Regional Development Fund (Photonics for Society P4S) and the CNRS (IRP LAFONI).

\bibliographystyle{apsrev4-1}
\bibliography{Refs}

\end{document}